\documentclass[twocolumn]{article}

\usepackage[modulo,switch]{lineno}
\modulolinenumbers[1]

\usepackage[utf8]{inputenc}
\usepackage[english]{babel}
\usepackage{amsmath}
\usepackage{amssymb}
\usepackage{graphicx}
\usepackage[outdir=./]{epstopdf}
\usepackage{url}
\usepackage{extarrows}

\makeatletter\@ifundefined{date}{}{\date{}}
\makeatother

\evensidemargin\oddsidemargin \hoffset-22.4mm \voffset-28.4mm
\begin{document}

\title{Simulation of detecting contact nonlinearity in carbon fibre polymer using ultrasonic nonlinear delayed time reversal}

\author{Martin Lints$^{1),2)}$, Andrus Salupere$^{1)}$, Serge Dos Santos$^{2)}$ \\
$^{1)}$ Tallinn University of Technology, Department of Cybernetics,\\
   Akadeemia tee 21, 12618 Tallinn, Estonia. martin.lints@cens.ioc.ee\\
$^{2)}$ INSA Centre Val de Loire, Blois Campus;\\ COMUE "L\'{e}onard de Vinci", U930 "Imagerie et Cerveau" Inserm,\\ 3 rue de la Chocolaterie, CS23410, 41034 Blois, France.}

\maketitle\thispagestyle{empty}

\begin{abstract}
A finite element method simulation of a carbon fibre reinforced polymer block is used to analyse the nonlinearities arising from a contacting delamination gap inside the material. The ultrasonic signal is amplified and nonlinearities are analysed by delayed Time Reversal -- Nonlinear Elastic Wave Spectroscopy signal processing method. This signal processing method allows to focus the wave energy onto the receiving transducer and to modify the focused wave shape, allowing to use several different methods, including pulse inversion, for detecting the nonlinear signature of the damage. It is found that the small crack with contacting acoustic nonlinearity produces a noticeable nonlinear signature when using pulse inversion signal processing, and even higher signature with delayed time reversal, without requiring any baseline information from an undamaged medium.
\end{abstract}

\section{Introduction}

In the past, the use of carbon fibre reinforced polymer (CFRP) has been limited to non-structural parts of high-tech aeronautical products. In recent times, due to the effort of weight reduction and product lifetime enhancement, the application areas of CFRP have widened to the load-bearing parts of the aeronautical, automotive and civil engineering products. Due to the increased demands on the strength of the CFRP products and possible complex failure mechanisms, the Non-Destructive Testing (NDT) methods of CFRP have been an important applied and academic problem. 

The complex failure mechanisms of CFRP include microcracking and delamination. Microcracking can occur at lower loads or due to aging and can be difficult to examine using ultrasonic NDT. With increased loading, the damage can evolve to delaminations, a very fine cracking between the layers of the CFRP. These damages are difficult to detect using ultrasonic methods due to their small thicknesses. The damage can exhibit itself as a contact acoustical nonlinearity~\cite{solodov2002can}. A statistical distribution of microcracks or delamination damage in the material could also be described by hysteresis in a continuum material model~\cite{AleshinAbeeleHyst2004,solodov2001PhysRev,Solodov_nonlinearultrasonic}. This can also be applicable for other materials than CFRP, for example biological tissues~\cite{DosSantosPrevorovsky2011,riviere2012time}. Nonlinear ultrasonic methods have been in development for detecting and localizing fatigue and micro-crack damage by their nonlinear effects~\cite{guo2012detection,kober2014theoretical}. The detection of harmonic overtones is one of the simplest measures of nonlinearities~\cite{blanloeuil2016closed}. Many nonlinear analysis methods not requiring filtering have been developed, for example scaling subtraction method~\cite{bruno2009analysis,scalerandi2008nonlinear} or pulse inversion (PI) with its generalizations~\cite{ESAMDosSantos2008,ciampa2012nonlinear}, and applications of time reversal using scattering as new sources. 

This paper proposes a delayed TR-NEWS signal processing method~\cite{Lints2016} for detecting the nonlinear signature of a single small crack in CFRP as contact acoustical nonlinearity. In the Finite Element Method (FEM) simulation, the CFRP is modelled as anisotropic, layered medium. The ultrasonic signal is focused by TR-NEWS to the region of the material with the defect. The nonlinear signature of the crack is analysed by PI and compared with the delayed TR-NEWS method, which allows to create arbitrary wave envelope at the focusing region of TR-NEWS. It is used here to create an interaction of waves near the damage. The signature of the damage appears as the nonlinear effect of the wave interaction on the contacting crack. This signal processing requires only one transmitting and one receiving transducer. The effectiveness of the delayed TR-NEWS method has been shown in the previous work by physical experiments and simulations in undamaged and linear materials~\cite{Lints2016}. In this paper, the FEM simulation model is advanced further by including absorbing boundary conditions and the contacting crack defect in the material.

\section{Mathematical and simulation model}
This section describes the simulation which is based on a physical experiment, and describes the differences and similarities between the simulation and the experiment. It shows some important points about the mathematical model, the delayed TR-NEWS signal processing and the FEM simulation. Detailed information about the derivation of mathematical and FEM model is available at~\cite{MLRR2017}.

\subsection{Mathematical model}\label{sec:matmodel}
\begin{figure}[!htpb]
\centering
\includegraphics[width = 0.3\textwidth]{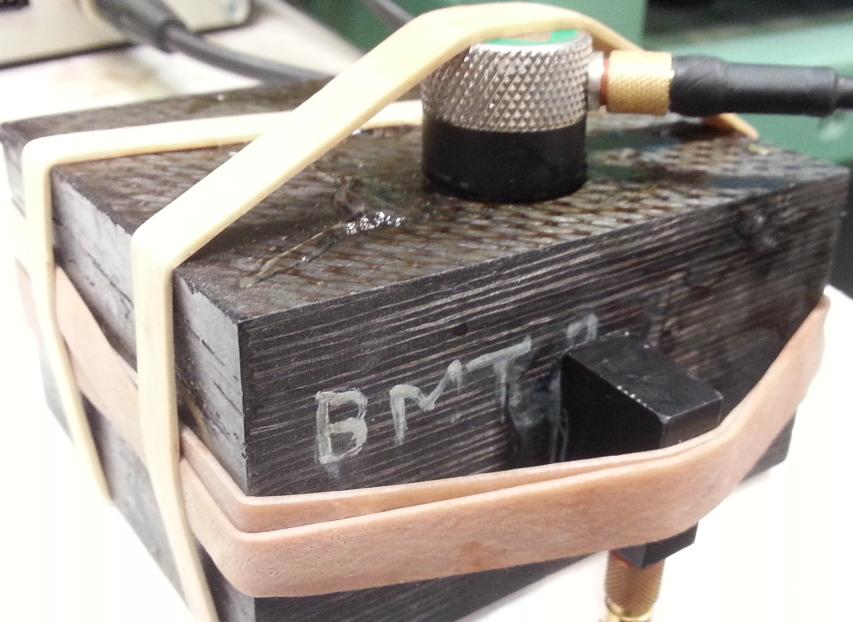}
\caption{(Colour online) CFRP block in the test configuration with transmitting transducer on side and receiving on top}
\label{fig:cfrptextconf}
\end{figure}

\begin{figure}[!htpb]
\centering
\includegraphics[width=45mm]{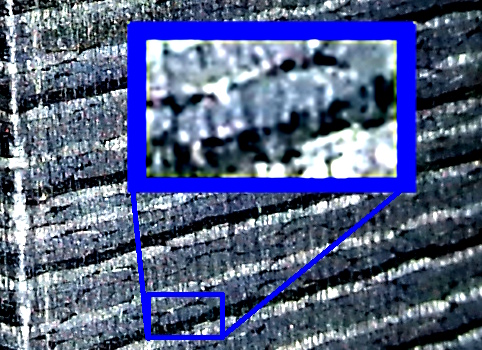}
\caption{The layered structure of the CFRP with the fabric yarns in tight packing and epoxy in the voids}
\label{fig:cfrpcloseup}
\end{figure}

The test object is a CFRP block consisting of 144 layers (Fig.~\ref{fig:cfrptextconf}).  It is composed of fabric woven from yarns of fibre and impregnated with epoxy. The cross-section of the yarns have elliptical shape (Fig.~\ref{fig:cfrpcloseup}) and the material has inclusions of pure epoxy, so a wave propagating through the material will encounter yarns (fibres with epoxy) and areas of pure epoxy. 

The simulation is in time domain, since the TR-NEWS procedure relies on transient echoes and complex wave motion for the wave energy focusing process. Due to the heavy computational cost of time domain simulation, a simple laminate model is used where: i)~the material consists of homogeneous layers, ii)~each layer has its own elasticity properties, and iii)~dispersion arises due to the periodical discontinuity of the material properties. It consists of CFRP layers with 90$^\circ$/0$^\circ$ weave, 45$^\circ$/45$^\circ$ weave and epoxy layer. The thicknesses of the layers are given by random variable functions which reflect the actual structure of the material. The random variable distribution, describing the CFRP structure, is measured from a close-up image of the CFRP test object~\cite{MLRR2017}. This links the distribution of the microstructure inside the actual material with the thicknesses of the layers in the laminate model. It should enable a more realistic simulated material having dispersion effects due to discontinuities.

The three different kind of layers have the following mechanical properties: i)~isotropic pure epoxy: $E = 3.7$~GPa, $\nu = 0.4$, $\rho = 1200$~kg/m$^3$; ii)~transversely isotropic composite with 0/90$^\circ$ weave: $E_1 = E_2 = 70$~GPa, $G_{12} = 5$~GPa, $\nu_{12} = 0.1$, $\rho = 1600$~kg/m$^3$; and iii)~ transversely isotropic composite with 45$^\circ$/45$^\circ$ weave: $E_1 = E_2 = 20$~GPa, $G_{12} = 30$~GPa, $\nu_{12} = 0.74$, $\rho = 1600$~kg/m$^3$. For the simulation, a laminate model was constructed using 50 pairs of epoxy and carbon fibre layers, where carbon fibre weave direction alternated between each pair. 

\begin{figure}[!htpb]
\centering
\includegraphics[width=45mm]{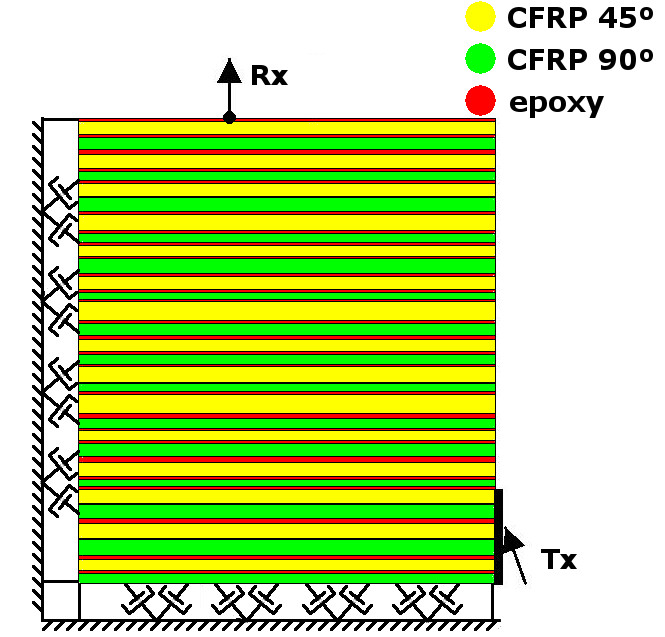}
\caption{(Colour online) The laminate material model with layers of stochastic thicknesses and absorbing boundary conditions on bottom and left boundaries and four fixed degrees of freedom}
\label{fig:laminatefig}
\end{figure}

\begin{figure}[!htb]
\centerline{\includegraphics[width=55mm]{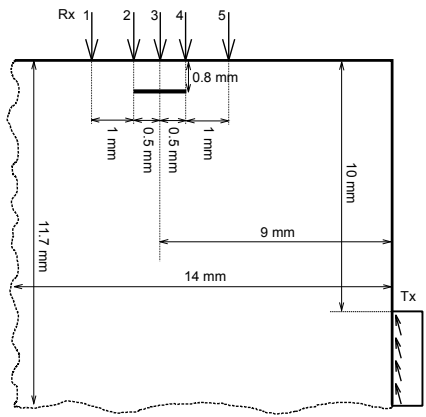}}
\caption{Schematic (not to scale) of the simulation geometry, location of crack, transmitter and receiver points  without the layers}
\label{fig:simsetup}
\end{figure}

The boundary conditions of the model (Fig.~\ref{fig:laminatefig}) include Lysmer-Kuhlemeyer absorbing boundary conditions~\cite{nielsen2006absorbing} so the wave energy would pass through the simulation region. Four degrees of freedom (DOFs) are fixed, the rest are free. The simulation model includes a contacting delamination defect in the material near the receiving transducer (Fig.~\ref{fig:simsetup}). The transmitting shear wave transducer can send maximum 50~kPa pulse at 70$^\circ$ degree angle.

\subsection{TR-NEWS signal processing}\label{sec:TRNEWS}
This subsection repeats the signal processing method that is applied for this problem and has been published in the previous work~\cite{Lints2016}. It is included for a self-contained discourse in this paper. 

In the physical experiments, on which the simulation is based on, the CFRP block (Fig.~\ref{fig:cfrptextconf}) was studied using TR-NEWS NDT equipment and signal processing methods~\cite{Lints2016}. The 2D FEM simulations reflect it as closely as possible in terms of transducer placement, frequencies and signal processing.

The roles of the transducers are not changed during the experiment: the focusing of the ultrasonic wave relies on the TR-NEWS signal processing. This is a two-pass method where the receiving and transmitting transducers do not change their roles. In this sense the ``Time Reversal'' describes the signal processing method which accounts for internal reflections of the material as virtual transducers, used for focusing the wave in the second pass of the wave transmission. The placement of the transducers is not important from the signal processing standpoint: in NDT investigation they could be placed arbitrarily and they do not have to be in line with each other, but the configuration must remain fixed during the complete TR-NEWS procedure.

\begin{figure}[htb]
\centering
\includegraphics[width=75mm]{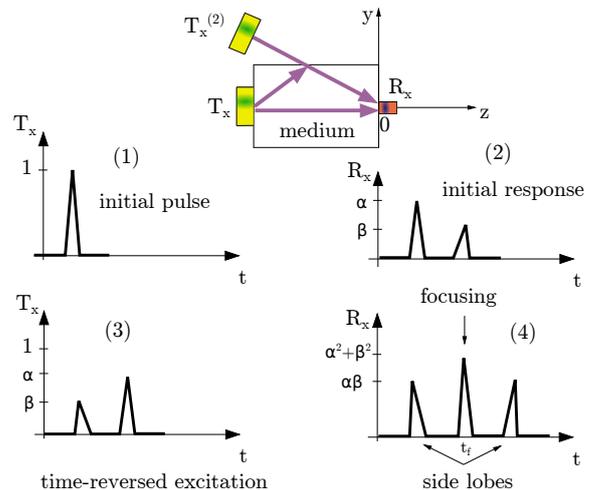}
\caption{(Colour online) Schematic process of TR-NEWS with the virtual transducer concept. (1) The initial  broadband excitation $T_x(t)$ propagates in a medium. (2) Additional echoes coming from interfaces and scatterers in its response $R_x$ could be associated to a virtual source $T_x^{(2)}$. (3) Applying reciprocity and TR process to $R_x$. (4) The time reversed new excitation $T_x = R_x(-t)$ produces a new response $R_x$ (the TR-NEWS coda $y_{TR}(t)$) with a spatio-temporal focusing at $z = 0; y=0; t=t_f$ and symmetric side lobes with respect to the focusing.}
\label{virttr}
\end{figure}

Figure~\ref{virttr} outlines the TR-NEWS signal processing steps. The simulation uses the same signal processing steps as are usually applied to physical experiments. Firstly the chirp-coded excitation $c(t)$ is transmitted through the medium.
\begin{equation}\label{eq:chirp}
c(t) = A \cdot \sin \left( \psi (t) \right),
\end{equation}
where $\psi(t)$ is linearly changing instantaneous phase. In this work, a linear sweep from 0 to 2 MHz was used. Then the chirp-coded coda response $y(t)$ with a time duration $T$ is recorded at the receiver
\begin{equation}
y(t,T) = h(t) \ast c(t) = \int_{\mathbb{R}} h(t - t',T) c(t') dt',
\end{equation}
where $h(t-t',T)$ is the impulse response of the medium. The $y(t,T)$ is the direct response from the receiving transducer when the chirp excitation $c(t)$ is transmitted through medium. Next the correlation $\Gamma(t)$ between the received response $y(t,T)$ and chirp-coded excitation $c(t)$ is computed during some time period $\Delta t$
\begin{equation}\label{gammaeq}
\Gamma (t) = \int_{\Delta t} y(t - t',T) c(t') dt' \simeq h(t) \ast c(t) \ast c(T-t,T),
\end{equation}
where the $h(t) \ast c(t) \ast c(T-t,T)$ is pseudo-impulse response which is proportional to the impulse response $h(t)$ if using linear chirp excitation for $c(t)$ because $ \Gamma_c (t) = c(t) \ast c(T-t) = \delta(t - T) $. Therefore the actual correlation $\Gamma (t) \sim h(t)$ contains information about the wave propagation paths in complex media. 

\begin{figure}[h!tb]
\centering
\includegraphics[width=75mm]{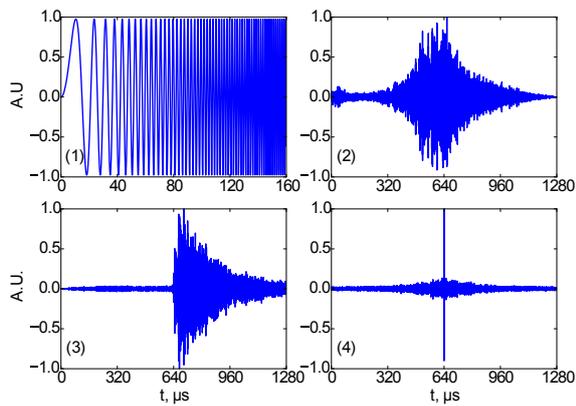}
\caption{Bi-layered aluminium experimental chirp-coded TR-NEWS signal processing steps: (1) chirp excitation, (2) output recorded at Rx, (3) cross-correlation between input and output, (4) focusing resulting from taking time-reversed cross-correlation as new input~\cite{Lints2016}.}
\label{steps4}
\end{figure}

Time reversing the correlation $\Gamma(t)$ from the previous step results in $\Gamma(-t)$ used as a new input signal. Re-propagating $\Gamma(-t)$ in the same configuration and direction as the initial chirp yields
\begin{equation}\label{resultequation}
y_{TR}(t, T) = \Gamma (T-t) \ast h(t) \sim \delta(t - T),
\end{equation}
where $y_{TR} \sim \delta(t - T)$ is now the focused signal under receiving transducer where the focusing takes place at time $T$. This is because $\Gamma (t)$ contains information about the internal reflections of the complex media, and transmitting its time reversed version $\Gamma (T-t)$ will eliminate these reflection delays by the time signal reaches the receiver, resulting in the focused signal $y_{TR}$ (Eq.~\eqref{resultequation}).
The test configuration must remain constant during all of these steps, otherwise the focusing is lost. The steps of this focusing process in a physical experiment are shown in Fig.~\ref{steps4}.

PI is an established method for detecting nonlinearities~\cite{ESAMDosSantos2008}. The procedure used here involves conducting TR-NEWS measurements with positive and negative sign for $A$ in Eq.~\eqref{eq:chirp} and comparing the focused signals. Differences could indicate the presence of nonlinearities.

\begin{figure}[h!tb]
\centering
\includegraphics[width=75mm]{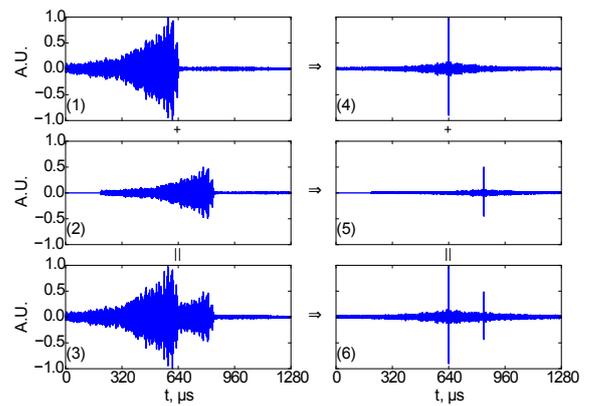}
\caption{Delayed TR-NEWS signal processing steps in bi-layered aluminium, starting from the cross-correlation step (left column) and prediction of linear superposition of waves (right column): (1) cross-correlation(Eq.~\eqref{gammaeq}), (2) delayed and scaled cross-correlation, (3) linear superposition of two cross-correlations which becomes the new excitation, (4) focusing (Eq.~\eqref{resultequation}), (5) delayed and scaled focusing, (6) linear superposition of the two focusing peaks.}
\label{steps6}
\end{figure}

Delayed TR-NEWS signal processing considers a single $y_{TR}$ focusing wave as a new basis which can be used to build arbitrary wave shapes at the focusing. This is done by time-delaying and superimposing $n$ time-reversed correlation $\Gamma(T-t)$ signals (Fig.~\ref{steps6} left column)
\begin{equation}\label{dTReq}
\Gamma_s(T-t) = \sum_{i=0}^n a_i \Gamma (T- t + \tau_i) = \sum_{i=0}^n a_i \Gamma (T-t + i \Delta \tau),
\end{equation}
where $a_i$ is the $i$-th amplitude coefficient and $\tau_i$ the $i$-th time delay. In case of uniform time delay the $\Delta \tau$ is the time delay between samples. Upon propagating this $\Gamma_s(t-T)$ through the media according to the last step of TR-NEWS, a delayed and scaled shape of signal at the focusing point can be created. The delayed TR-NEWS signal processing optimization can be used for amplitude modulation, signal improvement and sidelobe reduction~\cite{Lints2016}.

It is possible to predict what the delayed TR-NEWS focusing output would be in a linear material (Fig.~\ref{steps6} right column):
\begin{multline}\label{dTRpred}
y_{dTR}(t) = \left[ \sum_i a_i \Gamma_c \left( T - t + \tau_i  \right) \right] \ast h(t) \xlongequal{\text{linearity}} \\= \sum_i a_i \Gamma_c \left( T - t + \tau_i \right) \ast h(t) = \sum_i a_i y_{TR} (t - \tau_i).
\end{multline}
The purpose of the prediction is twofold. Firstly it can be used to figure out optimal delay and amplitude parameters $a_i$ and $\tau_i$ beforehand for the delayed TR-NEWS experiment, using the original focusing peak $y_{TR}$. Secondly it could be possible to analyse the differences between the measured delayed TR-NEWS result and its prediction, which acts as a baseline for comparison. The difference could indicate the magnitude of nonlinearity, because the prediction relies on the applicability of linear superposition and is found to be accurate in experiments with linear material~\cite{Lints2016}.

\subsection{The FEM simulation model}
The simulation program considers 2D wave propagation in a solid material with linear elasticity. The nonlinearity comes from an internal defect, a crack in the computational region (Fig.~\ref{fig:simsetup}) which can come into contact with itself. This contacting nonlinearity has asymmetric stiffness and is therefore nonclassically nonlinear. Since the CFRP is a complex material, then in this work it is modelled as a laminate with anisotropic layers arranged in a periodic manner, described in Section~\ref{sec:matmodel}. Because the physical experiment was conducted on the corner of a large CFRP block, the simulation is also in a semi-infinite quarter-space. The region has two free surfaces for reflection and two absorbing boundaries for the wave energy to escape.

The constitutive equation of the material itself is linear (although anisotropic). The linear plane strain elastodynamics problem is solved
\begin{equation}\label{eq:linelast}
\rho \ddot{u}_i - \sigma_{ij,j} = b_i,
\end{equation}
where $\rho$ is material density, $u_i$ is displacement component, $\sigma_{ij}$ is stress component and, $b_i$ is body force component~\cite{reddy}. Einstein summation convention is used and comma in index denotes spatial derivative. The constitutive equation in the variational formulation is 
\begin{multline}\label{eq:varform}
0 = \int_{\Omega} \left( \sigma_{ij} \delta \varepsilon_{ij} + \rho \ddot{u}_i \delta u_i \right) dx dy - \\ 
\int_{\Omega} b_i \delta u_i dx dy - \int_{\Gamma} t_i \delta u_i ds
\end{multline}
where $\varepsilon_{ij}$ is strain and $t_i$ is traction component on boundary. In our case the region $\Omega$ is a 2D space and boundary $\Gamma$ surrounding it is a 1D line. The body forces are zero in this simulation. Strain is assumed to be small. 

The matrix formulation of the finite  element model with damping is
\begin{equation}\label{eq:dynsystem1}
M\ddot{\Delta} + C\dot{\Delta} +K \Delta = F, 
\end{equation}
where $M$ is mass matrix, $K$ is stiffness matrix, $F$ is external forcing and $\Delta$ is displacement vector~\cite{MLRR2017}. The damping matrix $C$ is used to apply the Lysmer-Kuhlemeyer absorbing boundary conditions~\cite{nielsen2006absorbing} as a diagonal matrix, allowing to take advantage of the explicit solution scheme.

The element matrices are
\begin{equation}\label{eq:consistentmass}
M_e = \int_\Omega \rho \Psi^T \Psi dx dy,
\end{equation}
\begin{equation}
K_e = \int_\Omega B^T C_e B dx dy,
\end{equation}
\begin{equation}
F = \int_\Gamma \Psi^T f ds,
\end{equation}
where $C_e$ is here the constitutive matrix for the plane strain elasticity.

Linear triangular three-node elements (T3), also known as constant strain triangles~\cite{reddy}, were chosen for this problem for the following reasons. Firstly because the epoxy layers in the laminate model can be very small, therefore small elements are needed anyway, with T3 being computationally cheapest. Secondly, linear elements are well suited for nonlinear problems: since the strain is constant throughout the element, the computation of nonlinear constitutive relations would also be simple. In this simulation, the material itself is linear but future work might include nonlinearity or hysteresis. 

The T3 element lumped mass matrix~\cite{zienkiewicz1} is  
\begin{equation}
M_e = \frac{\rho A_e}{3} I_6,
\end{equation}
where $I_6$ is $6 \times 6$ identity matrix and $A_e$ is the area of the element.
The element stiffness matrix is 
\begin{equation}
K_e = A_e B_e^T C_e B_e,
\end{equation}
where matrix $B$ is
\begin{equation}\label{eq:bmatrixT3}
B = \frac{1}{2A_e}
\begin{bmatrix}
\beta_1 & 0 & \beta_2 & 0 & \beta_3 & 0 \\ 
0 & \gamma_1 & 0 & \gamma_2 & 0 & \gamma_3 \\ 
\gamma_1 & \beta_1 & \gamma_2 & \beta_2 & \gamma_3 & \beta_3 \\ 
\end{bmatrix},
\end{equation}
and with $x_i$ and $y_i$ being the node coordinates~\cite{reddy}, then
\begin{equation}
\begin{array}{llcrr}
\beta_1 &= y_2 - y_3,  & \quad & \gamma_1 &= x_3 - x_2, \\
\beta_2 &= y_3 - y_1,  & \quad & \gamma_2 &= x_1 - x_3, \\
\beta_3 &= y_1 - y_2,  & \quad & \gamma_3 &= x_2 - x_1. \\
\end{array}
\end{equation}

The external distributed force is simply divided into relevant nodes. The Lysmer-Kuhlemeyer absorbing boundary conditions are applied as viscous stresses on the boundaries, which means that they can be applied on DOF basis, making the damping matrix $C$ diagonal. The viscous stresses on the boundary DOFs are
\begin{align}
c_{ii} &= \int_\Gamma a \rho V_p ds, \quad \text{normal motion DOF},\\
c_{ii} &= \int_\Gamma b \rho V_s ds, \quad \text{shear motion DOF},
\end{align}
where $\Gamma$ is the boundary portion of the element~\cite{nielsen2006absorbing}. In this work the scaling parameters are $a=1$ and $b=1$. The wave velocities used for these boundary conditions are $V_p = 2972$~m/s and $V_s = 1956$~m/s~\cite{advancedNDT}.

Equation~\eqref{eq:dynsystem1} is solved for each timestep $\Delta t = 5\cdot 10^{-10}$~s by explicit central difference scheme
\begin{multline}\label{eq:lumpedsol}
\left( \frac{M}{\Delta t^2} + \frac{C}{2 \Delta t} \right) u_{n+1} = F_n -\\ \left( K - \frac{2 M}{\Delta t^2} \right) u_n - \left( \frac{M}{\Delta t^2} - \frac{C}{2 \Delta t} \right) u_{n-1}.
\end{multline}
This scheme is solved by dividing the equation by the term in the first parentheses, which is simple if $M$ and $C$ are diagonal matrices. Each simulation considers a 60~$\mu$s time window.

\subsubsection{Contact gap treatment}
There is a single source of nonlinearity in this simulation: the contacting crack fully inside the material~(Fig.~\ref{fig:simsetup}). If the material is at rest, then the crack is small and straight. In this work, there is neither a preload nor an initial gap in the contacting crack. This simple material defect results in a localised nonclassical nonlinearity, which can be analysed by various signal processing methods. 

It is known that frictional contact problems can be sensitive to timestep length and loading path~\cite{MijarArora2004}. In this work, it is assumed that the small timestep length and relatively small forces involved keep the error small. Therefore an explicit solution method scheme is utilized, similarly to \cite{SchutteEtAl2010}. A more precise solution could be expected from an implicit scheme, but that is left for the future. Further refinements could include thermoelastic contribution to the constitutive equation at the frictional contact gap~\cite{solodov2001PhysRev}.

The node-to-node contact model is used~\cite{zienkiewicz2} with Coulomb friction. The defect is horizontal, simplifying the calculation of normal gap between the nodes. If the position of a node on a slave (lower) surface is $(n^s_x, n^s_y)$ and on master (higher) $(n^m_x, n^m_y)$, then the normal contact gap is $g_N = n^s_y - n^m_y$ and the tangential gap (offset) is $g_T=n^s_x - n^m_x$. In case of normal penetration of one surface into another, then $g_N > 0$. If there is no penetration, then $g_N \leq 0$. The coefficient of friction is $\mu = 0.6$, and the solution aims to satisfy the Kuhn-Tucker conditions on the crack surface:
\begin{equation}
\begin{cases}
g_N \leq 0,\\
\lambda_N = \sigma \cdot n \leq 0,\\
g_N \cdot \lambda_N = 0,
\end{cases}
\end{equation}
where $\lambda_N$ is the normal force on crack, $\sigma$ is stress and $n$ is the normal vector of the surface. The penalty plus Lagrange multiplier method is used for normal contact and the penalty method for friction~\cite{MijarArora2004p1}.

The contact logic for the node pairs can be summarized by following steps.
\begin{itemize}
\item The initial contact forces are zeroed: normal $\lambda_N = 0$ and tangential $\lambda_T = 0$.
\item System in Eq.~\eqref{eq:lumpedsol} is solved.
\item Vector gap functions are found: $g_N = n^s_y - n^m_y$  and $g_T = n^s_x - n^m_x$.
\item Normal forcing is updated $\lambda_N = \lambda_N + g_N b$ where $b$ is some big penalty value and $\lambda_N \geq 0$.
\item Logic diverges to 3 paths:
\begin{description}
\item[No force is applied] in case of no contact.
\item[Only normal force] is applied if preceding step had no contact or had contact with tangential slip.
\item[Normal and tangential forces] are applied if previous iteration had non-slip contact.
\end{description}
\item The normal contact condition is verified by setting the penetration value $g_P = g_N$ where $g_N \geq 0$. Then the $L^2$-norm of penetration is evaluated $\langle g_P | g_P \rangle < \varepsilon$ where $\varepsilon$ is the limiting value for the error due to contact penetration. If the condition is not fulfilled, the iteration is repeated, otherwise new timestep is taken. 
\end{itemize}
A more thorough explanation of this contact gap logic is available at~\cite{MLRR2017}.

\section{Results}
The signal analysis of the time domain simulation results of the damaged and undamaged medium are compared, describing some analysis measures which could allow to detect the presence of damage as nonlinearity. The simulation follows ultrasonic TR-NEWS NDT procedures where the transducer data is available as time-series, measured at some specific location. The signals are low-pass filtered to keep only the ultrasonic component. Here five measurement points are analysed at various distances from the crack damage and transmitting transducer~(Fig.~\ref{fig:simsetup}). A video of the displacement fields for TR-NEWS focusing to point 3 in cracked medium is available at~\cite{CFTRfoc3video}. 

\subsection{TR-NEWS with pulse inversion}
\begin{figure}[!htb]
\centerline{\includegraphics[width=75mm]{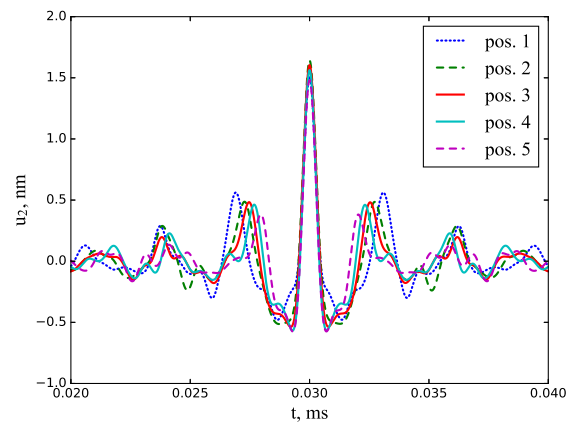}}
\caption{(Colour online) Unnormalized TR-NEWS focusing of undamaged CFRP simulation}
\label{fig:noncr}
\end{figure}

Figure~\ref{fig:noncr} shows the undamaged CFRP TR-NEWS focusing for the receiver positions 1 to 5 (Fig.~\ref{fig:simsetup}). It is an ordinary TR-NEWS focusing where at the middle of the signal (30~$\mu$s) is the focusing, surrounded by the sidelobes. There are two aspects to note about this is figure. Firstly, the sidelobes shift toward the main focusing and comparatively decrease in amplitude as the receiving transducer position shifts toward the transmitting transducer (from position 1 to position 5), indicating lower noise as the signal gets stronger. Secondly, the sidelobes are symmetrical in respect to the main lobe. This does not happen in nonlinear (damaged) material. The PI results are identical, indicating no nonlinearity, and are not shown here.

\begin{figure}[!htb]
\centerline{\includegraphics[width=75mm]{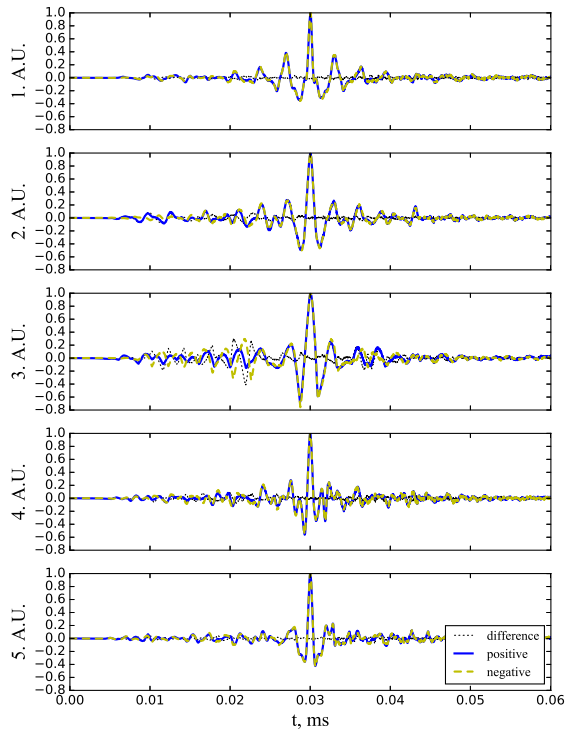}}
\caption{(Colour online) Normalized TR-NEWS focusing of damaged CFRP simulation with PI applied to detect nonlinearities as difference between negative and positive excitations}
\label{fig:cr5plot}
\end{figure}

Figure~\ref{fig:cr5plot} shows the TR-NEWS results of the cracked CFRP test object simulation for the receiving transducer positions 1 to 5 (Fig.~\ref{fig:simsetup}). Here the PI signal processing is also applied and it shows the nonlinearity as difference between results from initial chirp signals with positive and negative sign. This nonlinear, damaged case exhibits nonlinearity particularly strongly in receiving position 3 (near the middle of the crack). Also, the sidelobes are unsymmetrical in respect to the main lobe.

\begin{figure}[!htp]
\centerline{\includegraphics[width=75mm]{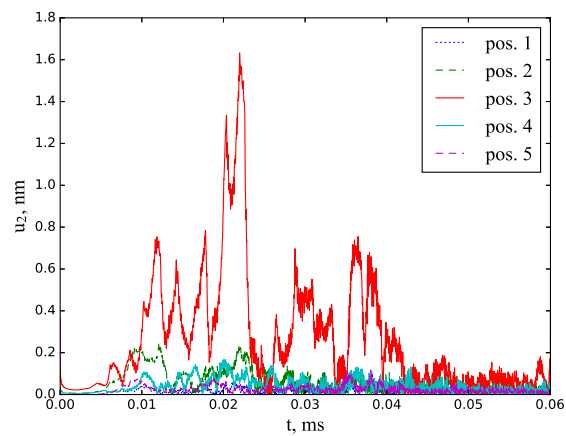}}
\caption{(Colour online) Envelopes of the PI nonlinearity measures from all of the measuring points}
\label{fig:nl}
\end{figure}

Figure~\ref{fig:nl} shows the envelopes of the PI measure of nonlinearity across the measuring points. The nonlinearity magnitude depends on the measuring point location in respect to the crack: point 3 near the middle of the crack shows strongest nonlinearity, points 2 and 4 show less, and points 1 and 5 show the least.

\begin{figure}
\centerline{\includegraphics[width=75mm]{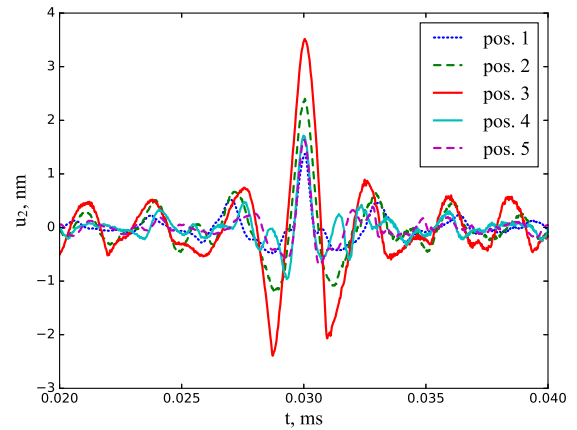}}
\caption{(Colour online) Unnormalized TR-NEWS focusing of damaged CFRP simulation}
\label{fig:cr}
\end{figure}

Figure~\ref{fig:cr} shows the unnormalized focusing signal for the damaged medium, which can be compared with corresponding undamaged result in Fig.~\ref{fig:noncr}. The focused signals have some interesting properties:
\begin{enumerate}
\item The highest signal amplitude comes from the receiver position closest to the crack midpoint (pos.~3), not from the position closest to the transmitter (pos.~5).
\item Comparing the amplitudes of the positions 2 and 4, at far and near side of the crack end respective to transmitter: the farther position has larger focusing amplitude than the nearer position. Since the simulation region has two absorbing boundaries, the wave propagation is mostly in one direction, therefore the defect between pos.~2 and~4 must be capturing the wave energy and the TR-NEWS signal processing is using that energy as a new ``virtual transducer'' for the pos.~2 focusing. This could be further analysed in future works from the correlation signals which generate these focused signals.
\item Amplitudes from the measurement positions~1 and~5 are ``right way'' around: the nearer measurement point has larger focusing amplitude than the farther. 
\end{enumerate}

\begin{figure}[!htb]
\centerline{\includegraphics[width=75mm]{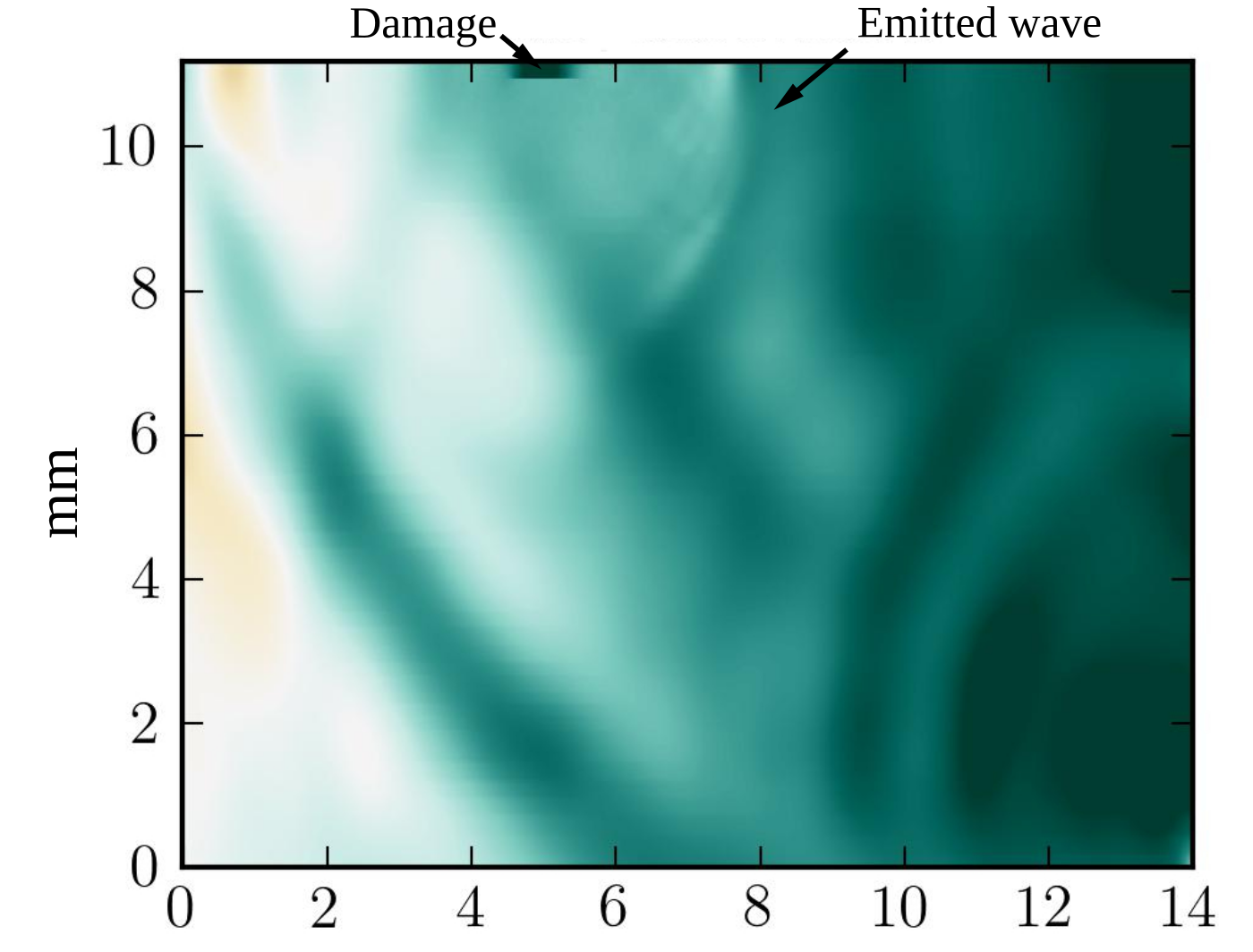}}
\caption{(Colour online) Displacement $u_2$ field at time $t=32.6$~$\mu$s with a wave emission coming from the damaged region. Video available at~\cite{CFTRfoc3video}}
\label{fig:virtsource}
\end{figure}

Figure~\ref{fig:virtsource} shows a snapshot of the simulation $u_2$ displacement at a time moment $t=32.6$~$\mu$s, just after the focusing. The defect in material is acting as a source of new excitation after TR-NEWS focusing. Wave energy is captured between the damage and outside wall of the material and emitted as a wave.

\subsection{Delayed TR-NEWS analysis}
\begin{figure}[!htp]
\centerline{\includegraphics[width=75mm]{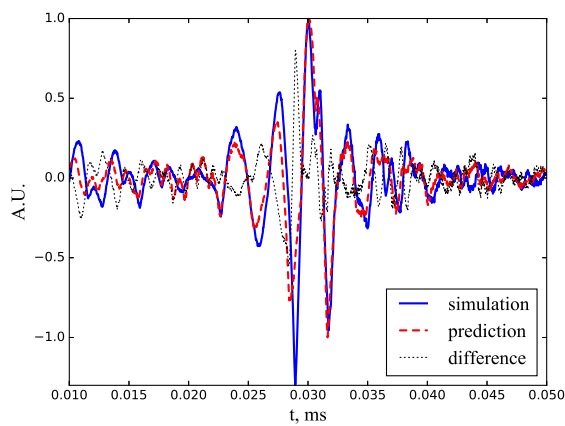}}
\caption{(Colour online) Delayed TR-NEWS signal processing with one delay of amplitude $a_i=1$ and delay value $\tau=1$~$\mu$s (Eq.~\eqref{dTRpred}): comparison between the linear prediction and the nonlinear simulation outcome}
\label{fig:dTR}
\end{figure}

Section~\ref{sec:TRNEWS} describes the delayed TR-NEWS signal processing method which allows to create arbitrary envelope wave at the focusing (Eq.~\eqref{dTReq}), instead of the simple peak of the TR-NEWS. Equation~\eqref{dTRpred} shows that in linear material, the outcome of the delayed TR-NEWS process can be predicted. Since this method with prediction works very well in physical NDT measurements of linear materials~\cite{Lints2016}, it is now tested in simulation with the nonlinearity, supposing that the difference between the simulation result and the linear prediction (Eq.~\eqref{dTRpred}) is due to nonlinear interaction of waves in the presence of nonlinearities or damage. Figure~\ref{fig:dTR} shows the comparison between the linear superposition prediction and the simulation result of a simple delayed TR-NEWS process where two focusing peaks are at superposition with 1~$\mu$s time delay. The difference between the prediction and the simulation is large and obvious, indicating the presence of nonlinearity. This measure of nonlinearity seems to be stronger than the measure calculated from PI (Fig.~\ref{fig:cr5plot}), making it a good candidate for further investigation.

The delayed TR-NEWS signal processing could also be used for activating the contacting gap as the energy pocket. This could be done by creating a new focusing wave envelope which would have the resonant frequency of the defect, permitting higher amplitude waves near the damaged region, which would enhance the extraction of the nonlinear signature. This study is left for the future.

\section{Conclusion}
This paper investigated nonlinear NDT by using a simple FEM simulation model for a crack nonlinearity in CFRP. In the laminate model, the damage is a simple horizontal contacting crack near the receiving transducer. The signal processing uses TR-NEWS method for focusing the available wave energy near the receiving transducer. The magnitude of nonlinearity due to the damage is measured firstly with PI, secondly with the proposed delayed TR-NEWS signal processing procedure. While PI indicates the presence of the nonlinearity, the simple delayed TR-NEWS procedure shows it even more strongly and is promising for future investigations and further development due to its signal processing flexibility.

Since the delayed TR-NEWS procedure allows to generate a wave at the focusing with arbitrary envelope, it could be used in the future to excite the crack damage by its resonance frequencies, using the damage as an energy pocket. Other perspectives include a more detailed simulation model for the CFRP in order to take more of its microstructure geometry into account to have stronger focusing. Additionally, the damage could be modelled either by a collection of various cracks at various angles or by hysteresis. Moreover, heating from the frictional forces at the damage could be considered for a more precise simulation model.

\subsection*{Acknowledgement}
This research has been conducted within the {\it co-tutelle} PhD studies of Martin Lints, between the Tallinn University of Technology, Department of Cybernetics in Estonia and the Institut National des Sciences Appliqu\'{e}es Centre Val de Loire at Blois, France. The research is supported by Estonian Research Council (project IUT33-24).

\end{document}